\documentclass[twocolumn,secnumarabic,amssymb, nobibnotes, aps, prd]{revtex4-1}
\usepackage{graphicx}
\usepackage{amsmath}
\usepackage{gensymb}
\usepackage{dcolumn}
\usepackage{bm}
\usepackage{amssymb}
\usepackage{color}
\usepackage[utf8]{inputenc}
\usepackage{url}

\begin{document}

\title{Wide tailorability of sound absorption using acoustic metamaterials}%

\author{A. Elayouch}%
\email[Corresponding author: ]{aliyasin.elayouch@femto-st.fr}
\author{M. Addouche}
\author{A. Khelif}
\affiliation{Institut FEMTO-ST, CNRS, Université de Bourgogne Franche-Comté, 15B avenue des Montboucons, 25030 Besancon Cedex, France}
\date{September 9, 2017}%

\begin{abstract}
	
	{\bf 
		
		We present an experimental demonstration of sound absorption tailorability, using acoustic metamaterials made of resonant cavities that does not rely on any dissipative material. As confirmed by numerical calculation, we particularly show that using quarter-wave-like resonators made of deep subwavelength slits allows a high confinement of the acoustic energy of an incident wave. This leads to enhance the dissipation in the cavities and, consequently, generates strong sound absorption, even over a wide frequency band. We finally demonstrate experimentally the key role of the filling ratio in tailoring such an absorption, using a metamaterial constituted of space-coiled cavities embedded in a polystyrene matrix. This paves the way for tremendous opportunities in soundproofing because of its low density, low volume, broadband and tailorable capabilities.

	} 
	
\end{abstract}

\maketitle

\section*{Introduction}\label{sec:intro}

The arrival of metamaterials came up with new designs allowing new wave manipulation functionalities, that might have been inaccessible with conventional materials due to their inherent limitations~\cite{cummer_controlling_2016}. Among them is the ability of generating broadband and omnidirectional absorption in a strong subwavelength regime which remains, more than ever, an important challenge~\cite{ma_acoustic_2016}. In acoustics, the paths leading to absorption are diverse, whether for ultrasound waves~\cite{leroy_superabsorption_2015} or sound~\cite{ma_acoustic_2014}. They are mainly based on the idea of converting the energy carried by an incident wave, whether through the conversion to another mode of vibration/propagation, or through the conversion of acoustic energy into heat. In this context, local resonances have showed their potential for generating remarkable sound opacity and transparency phenomena~\cite{christensen_theory_2008}, and when combined with viscoelastic or porous materials, astonishing sound absorption effects~\cite{ivansson_sound_2006}. For instance, a structure constituted of a membrane on which are deposited pairs of asymetric and rigid platelets can efficiently generate absorption phenomena at frequencies of resonance~\cite{mei_dark_2012}. Porous lamella-crystals backwarded with a reflector~\cite{christensen_extraordinary_2014} have moreover been proposed. It was particularly demonstrated a quasi-omnidirectionnal absorption for a frequency range exceeding two octaves. Other studies also highlighted that using one or two layers of resonant inclusions in a foam matrix, the all backed up by a rigid plate acting as a reflector, permits to reach high absorption at resonance frequencies of the inclusions~\cite{lagarrigue_design_2016}. 
However, more recently, acoustic metamaterials based on quarter-wave-like resonances have received particular attention, for several reasons. One of them is the fact that such metamaterials, unlike the above examples, does not rely on the use of materials having intrinsic dissipative properties with regard to the sound~\cite{Jiang_ultra-broadband_2014}. Indeed, it was notably demonstrated that combining a slowdown phenomenon of wave propagation with inherent dissipation can lead to efficient sound absorption effects~\cite{groby_use_2015, jimenez_quasiperfect_2017, santillan_subwavelength-sized_2016}. 
Furthermore, following on space-coiling technique recently introduced in acoustic metamaterials~\cite{liang_extreme_2012,frenzel_three-dimensional_2013, liang_space-coiling_2013, xie_measurement_2013, xie_tapered_2013,song_emission_2014}, it was demonstrated that low-frequency sound absorption can be achieved using coplanar spiral tubes acting as quarter-wave resonators~\cite{Cai_ultrathin_2014, zhang_three-dimensional_2016, li_acoustic_2016}. This permits to efficiently generate high level of sound absorption at a deep sub-wavelength regime. 
Indeed, when dealing with such subwavelength structures, it was notably showed that boundary layer effects cannot be neglected, even if they form only a tiny fraction of the cavities under consideration~\cite{ward_boundary-layer_2015}. From there, others structures consisting in deep subwavelength channel have been proven to be very efficient sound absorbing materials~\cite{moleron_visco-thermal_2016, starkey_thin_2017}. While various types of absorbers are presented with backward reflectors, ignoring the transmission properties that can arise from these acoustic structures, it could be very interesting to understand more precisely to what extent an absorber based on local resonances can control the acoustic properties. 

In this paper, we experimentally and numerically study the sound absorption capabilities of acoustic metamaterials made of resonant cavities, namely by considering acoustic metamaterials that combine both quarter-wave-like resonators and Fabry-Perot cavities. In doing so, we measure acoustic properties including transmission, reflection and absorption coefficients. Thus, such an acoustic absorber does not rely on any intrinsic dissipative material, which is mainly due to the strong subwavelength character of the geometries under consideration. We particularly investigate the confinement of acoustic energy in the cavities through the role of geometrical parameters, and its effect on the dissipation of sound. Based on this concept, we furthermore present an acoustic metamaterial that allows the generation of broadband sound absorption, and show the interaction that may be generated by such a combination, as well as its implications on transmission properties. Finally, we demonstrate through experimentation how sound absorption can be efficiently tailored using a deep subwavelength metamaterial made of space-coiled cavities.

\section{Energy Confinement and Dissipation Enhancement}

In this context, we may recall that absorption phenomena naturally exist during the propagation of sound waves. If viscous effects can generally be neglected for sound propagation  in free space, the situation is different when the dimensions of a structure become significantly smaller than the wavelength at the operating frequency, and the so-called shear viscosity can therefore no longer be overlooked. Indeed, Tijeman~\cite{tijdeman_propagation_1975} have highlighted the role of thermo-viscous effects on energy dissipation for the case of acoustic waves propagating in narrow pipes. Later, Wegdam et al. studied wave propagation in a crystal constituted of a fluid matrix whose viscosity directly affects the band gap characteristics~\cite{sprik_acoustic_1998}. Thus, viscous boundary layers $\delta_{v} = \sqrt{\frac{\eta}{\rho w}}$, with $ \eta $ the shear viscosity, $ \rho $ the fluid density, $w$ the angular frequency, and thermal boundary layers $\delta_{t} = \sqrt{\frac{\kappa}{\rho C_{p} w}}$ near the walls, with $\kappa$ the thermal conductivity, and $C_{p}$ heat capacity at constant pressure, are indicators which allow identifying whether thermo-viscous effects are likely to be involved or not, by comparing them to dimensions of the structure. For example, in the case of resonant transmission through a slit array, i.e. Fabry-Perot cavities, formed of aluminum slats, Ward \emph{et al.} highlighted a significant 5\% reduction of the effective speed of sound with a boundary layer occupying only 5\% of the total slit width, resulting in a substantial damping of the resonance as well as a shifting of the resonant frequency~\cite{ward_boundary-layer_2015}.

\begin{figure}[!h]
	\centerline{\includegraphics[width=8.6cm,keepaspectratio]{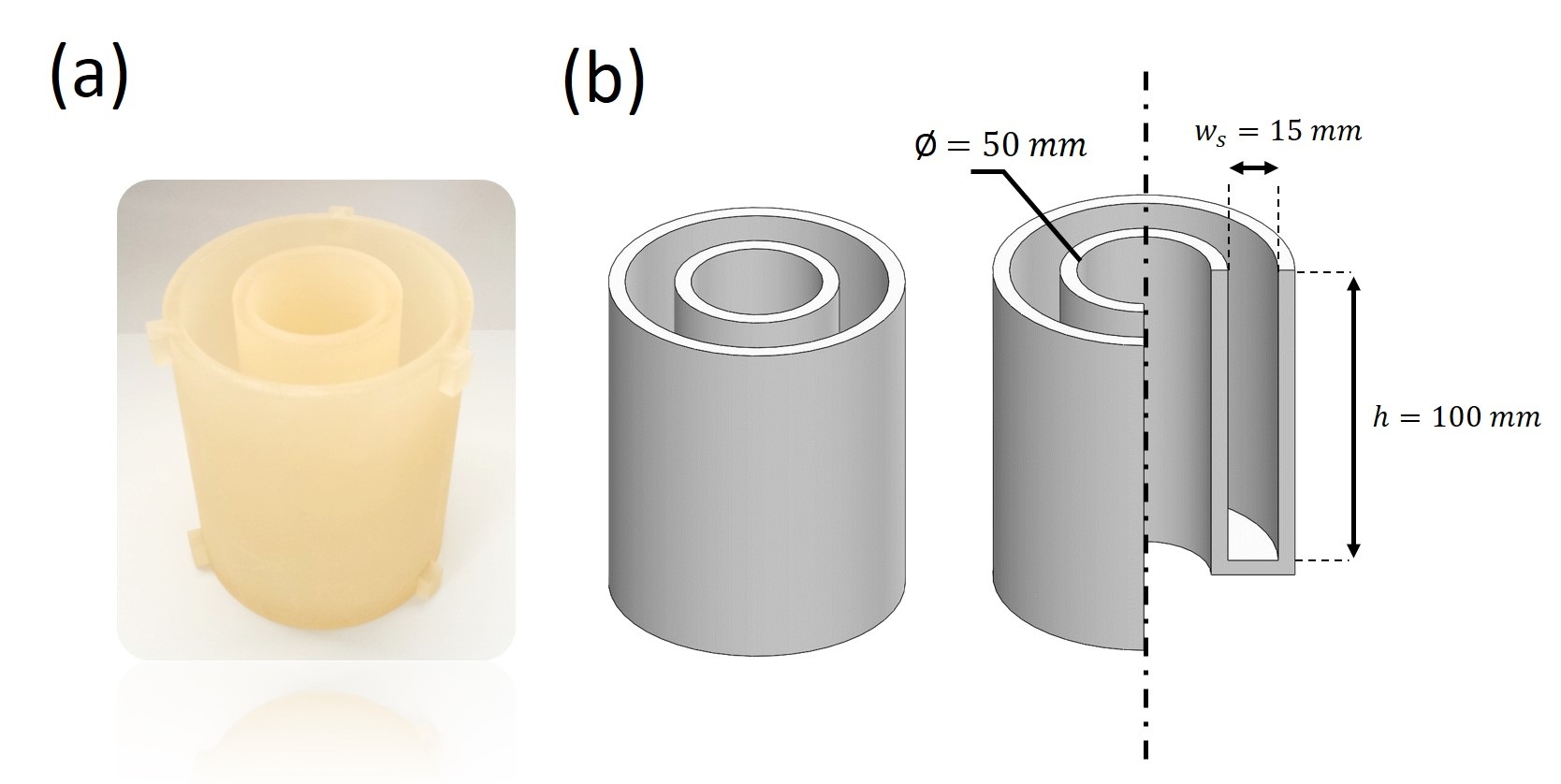}}
	\caption{(a) Photograph of the acoustic metamaterial fabricated by 3D printing. The sample is constituted by a quarter-wave-like resonator taking the shape of a hollow cylindrical pipe - circular slit -, closed-ended at one side. 
		(b) Schematics of the acoustic metamaterial fabricated by 3D printing. The object is cut using two imaginary cutting planes. The unwanted portion is mentally discarded exposing the interior structuration. The width of the slit equals $w_s$. There is an interior and exterior radius of $r_i=25$ mm and $r_e = r_i+w_s$. The walls have a thickness of $t=5$ mm. The metamaterial height is equal to $h=100$ mm.}
	\label{fig:fig01}
\end{figure}
In our work, we consider the combination of a quarter-wave-like acoustic resonator, constructed as a hollow cylindrical pipe, closed-ended at one side, surrounded by a Fabry-Perot cavity as presented in Figure~\ref{fig:fig01}. Regarding the quarter-wave-like resonator, the resonance frequency of the fundamental mode depends on the length of the $h$ pipe, as $f=w/2 \pi=c/4h$, with $c$ the sound velocity in air. The interior radius of the cavity and the width of the slit, identified as the parameters $r_i$ and $w_s$ permit, among others, to control the quality factor of the resonance (see Figure~\ref{fig:fig01}b). We are interested in obtaining the acoustic transmission properties of the metamaterial, at normal incidence. The experimental measurements are realized using a Kundt's tube, which is basically a standing wave tube. The inner diameter of the tube section is 100 mm. A loudspeaker, at one end of the tube, is used for generating a broadband random signal over the frequency range 50–1500 Hz. Acoustic pressure measurements are carried out for two different tube termination conditions, open and approximately anechoic.

\begin{figure}[!h]
	\centerline{\includegraphics[width=8.6cm,keepaspectratio]{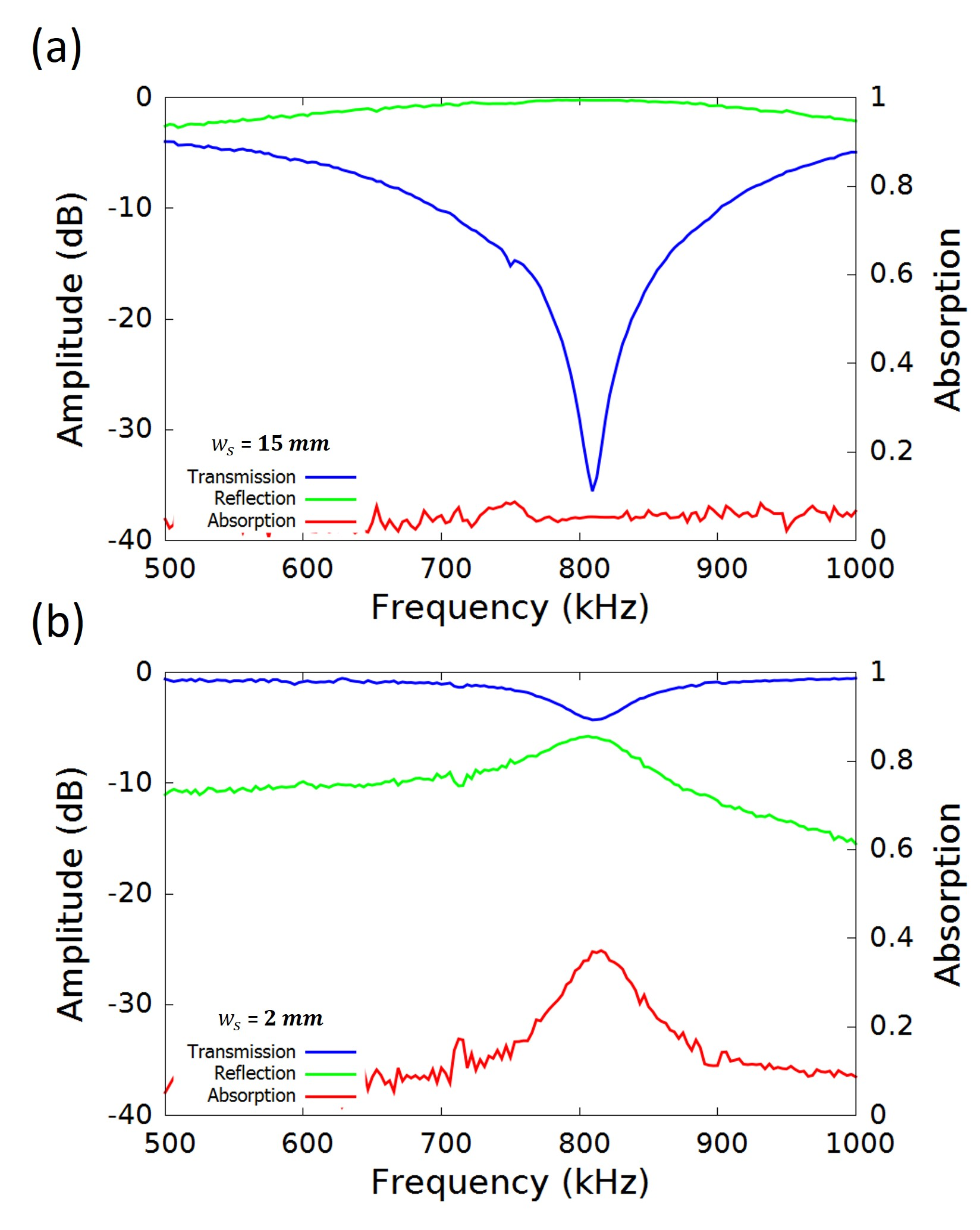}}
	\caption{Amplitude of the transmission (blue), reflection (green) and absorption (red) spectra obtained experimentally. The results are related to the sample made of a hollow cylindrical pipe - circular slit -, closed-ended at one side, with the following geometrical parameters: $h=100$ mm, $r_i=25$, and $w_s=15$ mm.}
	\label{fig:fig02}
\end{figure}

In the following, we first experimentally study the influence of the slit width $w_s$ on the variation of the absorption. The dimensions of the acoustic metamaterial are chosen to position the first resonance mode under the frequency of $1$ kHz. The proposed structures are “fabricated” using the 3D printer Projet SD3500 (see Figure~\ref{fig:fig01}a). They are held, at three connection points located at the upper and lower sections of the pipe, by a cylindrical support having an external radius equal to the radius of the Kundt’s tube and a thickness of $3$ mm. A three-dimensional representation of the metamaterial is shown in Figure~\ref{fig:fig01}b. In the present case, the pipe have a length of $h=100$ mm, and a radius of $r_i=25$ mm, and a slit width of $w_s=15$ mm. We obtain the following transmission, reflection and absorption spectra, presented in Figure~\ref{fig:fig02}a. We clearly observe the typical behavior of a quarter-wave acoustic resonator, that generates a near-total reflection at the resonance frequency. The amplitude of transmission reaches  $-35dB$, and a reflection coefficient of nearly $0$ dB at the frequency of $810$ Hz. Moreover, it can be seen that the coefficient of absorption does not exceed $5\%$. When we change the slit width from $w_s=15$ mm to $w_s=2$ mm, we can see that the transmission dip is greatly reduced from $-35dB$ to nearly $-5dB$ (see Figure~\ref{fig:fig02}b), and the absorption coefficient increased, reaching almost $40\%$. This indicates that much more energy is dissipated, which affects the quality factor of the resonance. Besides, since the effective area of the Fabry-Perot cavity is increased, there is globally more transmission all over the frequency range under consideration. Indeed, the reflection coefficient is between $-15$dB and $-7$dB for $w_s=2$ mm. 

\begin{figure}[!h]
	\centerline{\includegraphics[width=8.6cm,keepaspectratio]{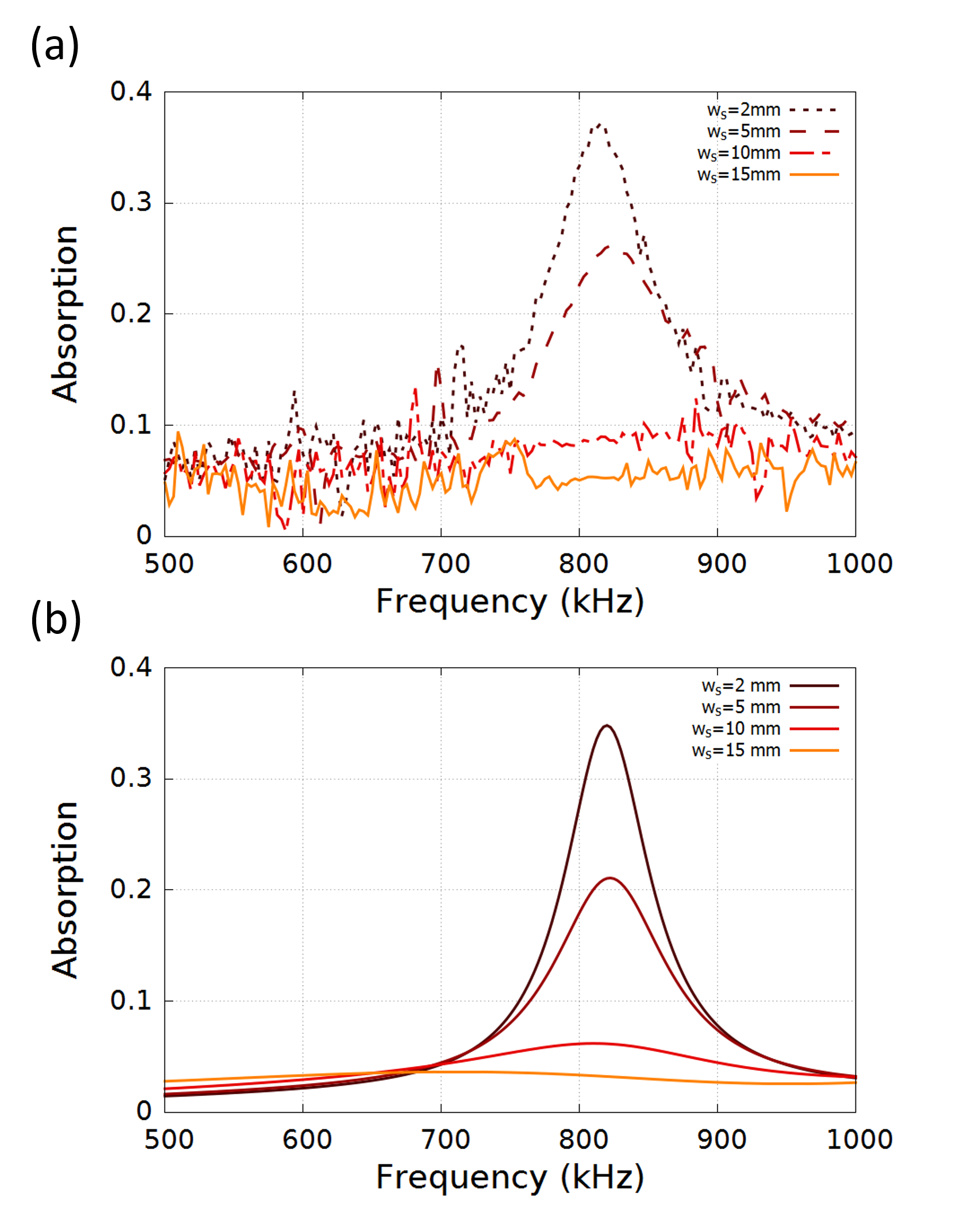}}
	\caption{Comparison of the absorption spectra related to the hollow cylindrical pipe - circular slit -, closed-ended at one side, with the following geometrical parameters: $h=100$ mm, $r_i=25$, but for different configurations of slit width $w_s=15, 10, 5$ and $2$ mm. (a) is obtained experimentally using a Kundt's tube and (b) represents the spectra obtained numerically. }
	\label{fig:fig03}
\end{figure}

We can now compare, in Figure~\ref{fig:fig03}a, the absorption spectrum obtained experimentally for samples having different values of slit width $w_s=15, 10, 5$ and $2$ mm. A first and obvious observation is the local rising of the absorption coefficient from $5$, $8$, $26$ to $37\%$, as the slit width parameter decreases. At this operating frequency of $810$ Hz, the wavelength is $42,3$ cm, which is nearly two hundred times larger than a slit width of $w_s=2$ mm. The boundary layers represent $\frac{\delta_{v} }{w_s} = 2.8\%$ and  $\frac{\delta_{t} }{w_s} = 3.2\%$ of the slit width. Besides, the filling ratio of the quarter-wave-like resonator, i.e. effective area of the cavity, is function of $w_s$ as $(\pi w_s(w_s+50))$. We reach an absorption coefficient of from $5\%$ to almost $40\%$, with a filling ratio roughly ten times reduced within a range from $40\%$ to $4\%$. It is worth highlighting that a similar observation have been made for the case of porous lamella-crystals~\cite{christensen_extraordinary_2014}.

To gain a better understanding of the mechanisms underlying the results presented above, we carry out a numerical study realized by performing finite elements simulations with Comsol, and take into account the dissipation phenomena previously mentioned. In this numerical analysis, one may use Kirchhoff theory that deals with total pressure $P$, total temperature $T$, total density $\rho$ and particle velocity $\vec{v}$, through the linearized Navier-Stokes equations, the continuity equation, the heat equation and the ideal gas law:

\begin{equation}
\rho_{0} \frac{\partial \vec{v}}{\partial t} = -\nabla P \frac{1}{3} \eta \nabla \left( \nabla \cdot \vec{v}\right) + \eta \Delta \vec{v} \,
\label{eq:eq42}
\end{equation} 

\begin{equation}
\rho_{0}  \nabla \cdot \vec{v} +  \frac{\partial \rho}{\partial t} = 0 \,
\label{eq:eq43}
\end{equation} 

\begin{equation}
\rho_{0} C_{p} \frac{\partial T}{\partial t} =  \kappa \Delta T + \frac{\partial P}{\partial t}  \,
\label{eq:eq44}
\end{equation} 

\begin{equation}
P=\rho R_{0} T \,
\label{eq:eq45}
\end{equation} 

with $ \rho_{0} $ is the background density of air and $ R_{0} $ specific gas constant of air. Then, it is a matter of linearizing these equations, and considering small harmonic perturbations around mean values. We model our structure, by taking advantage of the axial symmetry of the structure. We particularly consider rigid conditions at the boundaries of the solid domain, and we study the propagation of acoustic waves in air.
An acoustic source  is positioned on one side of the structure and we calculate the transmitted and reflected pressure field at both sides, which permits us to ultimately obtain the absorption spectrum for frequencies of interest to us. Moreover, we use Perfectly Matched Layers (PML) at the ends of the waveguide in order to avoid reflections at the limits, and thus create the virtual conditions of an infinite domain.

\begin{figure}[!h]
	\centerline{\includegraphics[width=8.6cm,keepaspectratio]{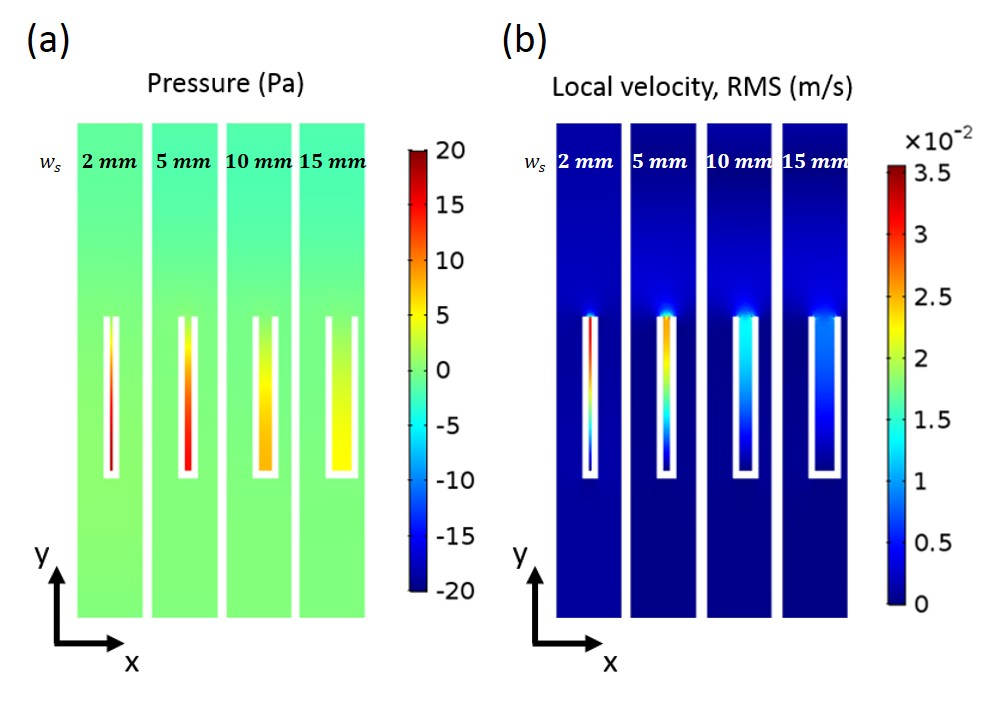}} 
	\caption{(a) Pressure fields and (b) fields of particle velocity vectors, in the quarter-wave-like cavity and in the surrounding Fabry-Perot cavity, at the resonance frequencies obtained numerically for different configurations of slit width $w_s=15, 10, 5$ and $2$ mm.}
	\label{fig:fig04}
\end{figure}

Figure~\ref{fig:fig03}b, that represents the corresponding absorption spectra, confirms the trend experimentally observed. The coefficient of absorption increases with the reduction of the slit width, at the resonance frequency of the cavity. In order to better visualize the reasons of this trend, we represent in Figure~\ref{fig:fig04} the pressure field at resonance, as well as the field of particle velocity vectors, for different values of $w_s$. We note that the pressure field has a very high amplitude in the narrow cavity, in comparison with the others area of the metamaterial,  as observed in Figure~\ref{fig:fig04}a. This comment applies equally to the amplitude of the field of particle velocity vectors which drastically increases in the narrow cavity, at the resonance frequency.

\begin{figure}[!h]
	\centerline{\includegraphics[width=8.6cm,keepaspectratio]{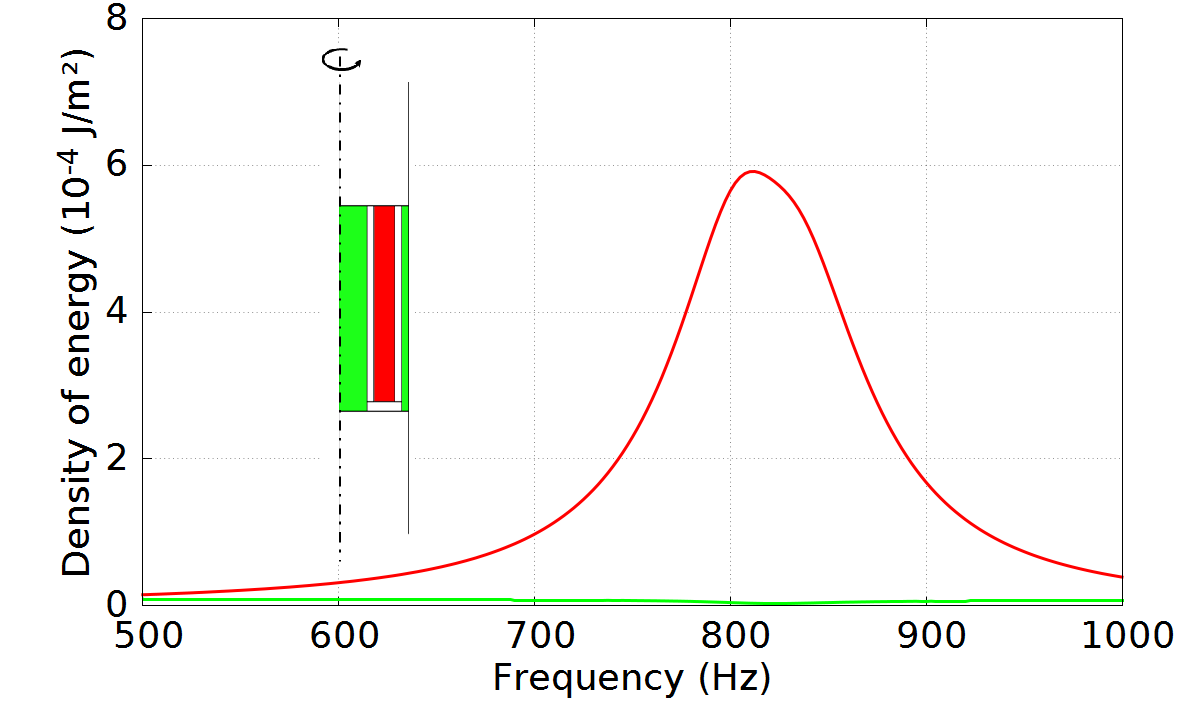}}
	\caption{Density of energy as function of frequency, calculated with the following geometrical parameters, $h=100$ mm, $r_i=25$, and $w_s=15$ mm, in the quarter-wave-like cavity (red curve), and in the surrounding Fabry-Perot cavity (green curve).}
	\label{fig:fig05}
\end{figure}

To illustrate the role played by the variables of acoustic pressure and particle velocities, we firstly represent in Figure~\ref{fig:fig05} the spectra of density of energy in the quarter-wave-like cavity, and in the surrounding Fabry-Perot cavity, for the configuration $w_s=15$ mm. This density of energy is calculated from the kinetic $\frac{1}{2} \rho |v|^2$ and potential energies $\frac{1}{2 \rho c^2} |p|^2$.
We can clearly observe how the energy stored in the quarter-wave-like cavity is much greater than in the surrounding Fabry-Perot cavity. We particularly reach a density of energy of $6.10^{-4} J/m^{2}$, which is $60$ times higher than the surrounding domain. 

\begin{figure}[!h]
	\centerline{\includegraphics[width=8.6cm,keepaspectratio]{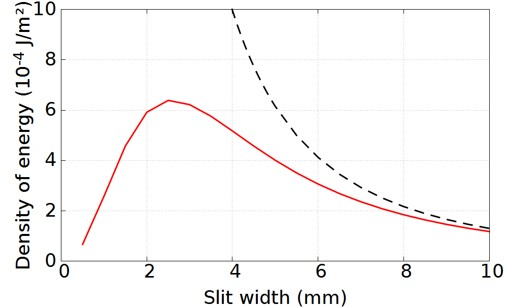}}
	\caption{ (a) Maximum density of energy reached and (b) its corresponding frequency, as a function of the slit width $w_s$ from $w_s=0.1$ mm to $w_s=10$ mm. The results are obtained numerically by finite element method, when considering the losses (red curve) and without losses (black curve).}
	\label{fig:fig06}
\end{figure}

Moreover, we represent in Figure~\ref{fig:fig06} the influence of the slit width on the energy density reached at the resonance frequency. We compare here the variation of energy density in loss-free and loss-inclusive models, in order to see how viscosity can affect the confinement of acoustic energy. In linear acoustics, i.e. loss-free model, the density of energy exponentially increases when decreasing the slit width parameter, as observed in Figure~\ref{fig:fig06}a. However, when thermal and viscous losses are introduced through the linearized Navier-Stokes equations, we observe that the density of energy confined in the quarter-wave-like cavity substantially increases with decreasing the slit width parameter until reaching a maximal value at $w_s=2.5$ mm. Thus, the slit width permits to control the level of acoustic energy confinement in the cavity. Such a confinement induces strong variations of the acoustic pressure and particle velocity fields, which are direcly related to visco-thermal effects and lead to enhancement of dissipation. The result is a significant sound absorption at the resonance frequencies of the cavity involved.

\section{Broadening and tailoring absorption}

\begin{figure}[!h]
	\centerline{\includegraphics[width=8.6cm,keepaspectratio]{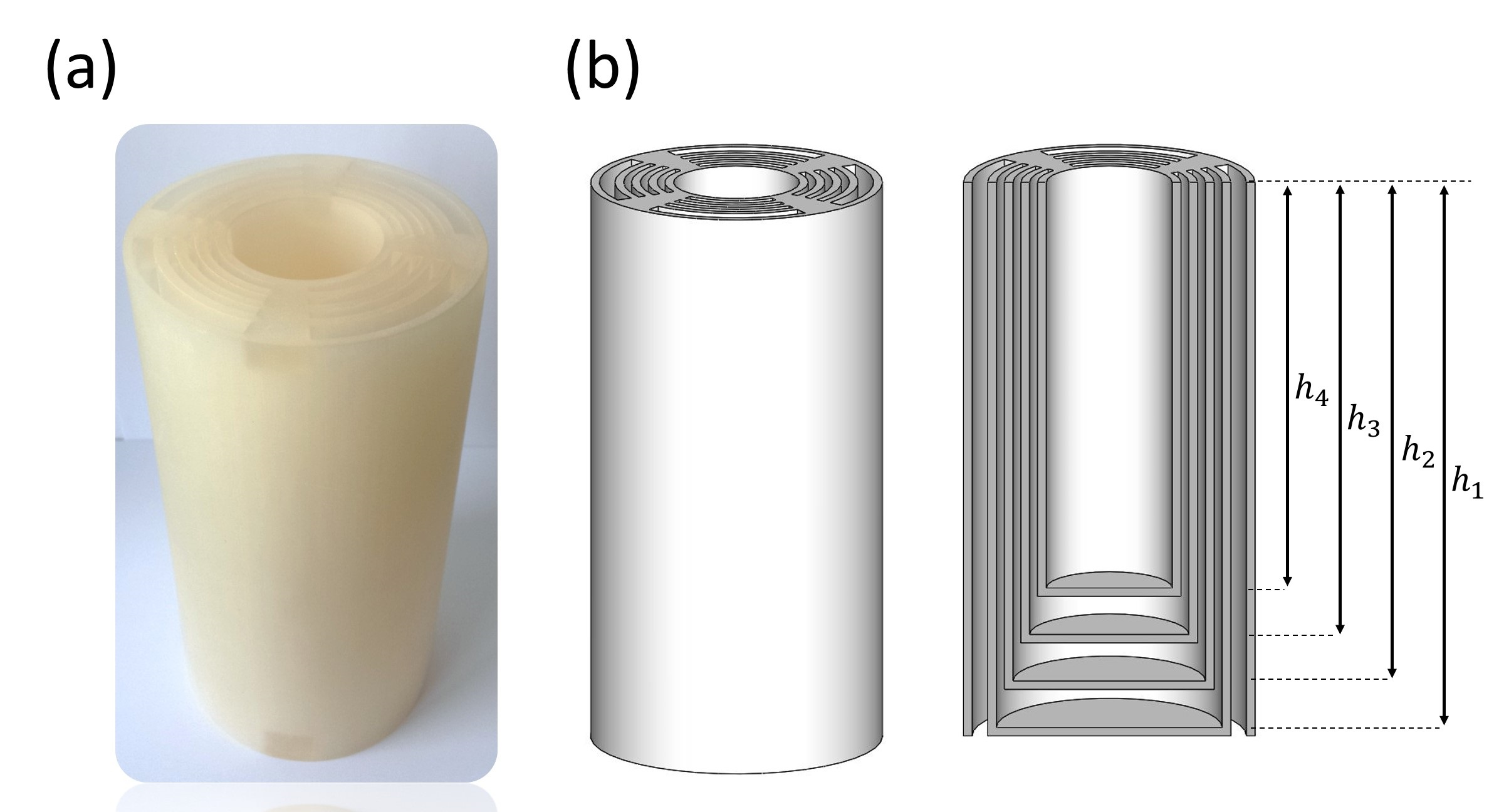}}
	\caption{(a) Photograph of the an acoustic metamaterial constituted of multiple and neighboring resonant cavities, and fabricated by 3D printing. (b) Schematics of the acoustic metamaterial constituted of multiple and neighboring resonant cavities. The object is cut using an imaginary cutting plane. The unwanted portion is mentally discarded exposing the interior structuring. The metamaterial height is equal to $h=100$ mm.}
	\label{fig:fig07}
\end{figure}

In light of the above, we have concluded that a narrow quarter-wave-like cavity can generate a high level of acoustic absorption at the resonance frequency. The key question now is how to broaden the frequency range of acoustic absorption. The answer to this issue may lie in the possibility of using an acoustic metamaterial made of a series of neighboring resonant cavities in order to cover a large band of absorption. Such an acoustic metamaterial can take the form of more than one shorter and narrower hollow core tubes placed inside the upper most tube. An example of such a structure is represented in Figure~\ref{fig:fig07}, which consists of four end-capped pipes that fit into each other. The smallest pipes are “suspended” inside the closest wider one, using two connection points located at the upper sections of the pipes. The dimensions of the structure are: $(h_1,r_1) = (193.5,38.7)$ mm, $(h_2,r_2) = (177,33)$ mm, $(h_3,r_3) = (160.5,27.3)$ mm, $(h_4,r_4) = (144,21.6)$ mm, with $h_i$ and $r_i$ respectively the height and the radius of each end-capped pipe. All is held together by a cylindrical support having an external radius equal to the radius of the Kundt’s tube, and a height of $196.5$mm.  The thickness of walls equals $t=3$ mm. 

\begin{figure}[!h]
	\centerline{\includegraphics[width=8.6cm,keepaspectratio]{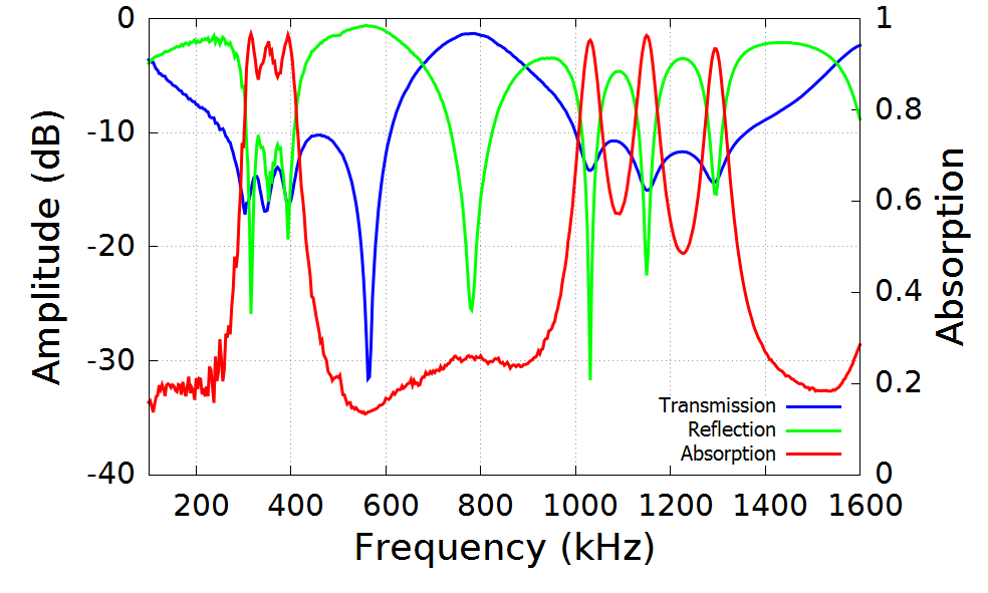}}
	\caption{Amplitude of the transmission (blue), reflection (green) and absorption (red) spectra obtained experimentally. The results are related to the acoustic metamaterial constituted of multiple and neighboring resonant cavities, represented in the frequency range from $100$ to $1600$ Hz.}
	\label{fig:fig08}
\end{figure}

We represent in Figure~\ref{fig:fig08} the transmission, reflection and absorption spectra obtained experimentally using the Kundt's tube. We clearly observe three peaks of absorption at the frequencies indicated in Table~\ref{tabchar_resultsM4}, with a maximal absorption of $0. 97$ reached at the frequency of $303$ Hz. These resonance frequencies, corresponding to the contribution of every cavity, are close to each other, which permits us to generate a wide frequency band of absorption centered at the frequency of $350$ Hz, that reaches $0.86$ on a relative bandwidth of $48.9 \%$. 
Moreover, we note that the first three frequencies of absorption are slightly shifted from the the dips of transmission, which does not correspond to what we observed in the case of only one resonator.  Indeed, this is likely due to the coupling between neighbor resonators that generates a series of asymmetrical profile of the transmission, and so-called Fano interaction~\cite{amin_acoustically_2015}. Thus, while the absorption phenomena is directly related to the high level of energy confinement in the cavity, the profile of the transmission is rather modulated by such an interaction.   
Finally, we can also remark that not only the fundamental modes, but also the harmonics are the place of absorption generation. Indeed, for these frequencies, the wavelength are far greater than the slits width, which consequently leads to strong dissipation of the acoustic energy. However, the harmonics are more-widely spaced, which diminishes the Fano interaction and  makes the transmission dips and absorption peaks coincide, as clearly observed in Table~\ref{tabchar_resultsM4}.

\begin{table}[!h]
	\caption{\label{tabchar_resultsM4} Absorption and transmission results}
	\centering
	{\renewcommand{\arraystretch}{1} 
		{\setlength{\tabcolsep}{0.18 cm} 
			\begin{tabular}{|c|c||c|c|}
				\hline
				\multicolumn{2}{c|}{Transmission dips} & \multicolumn{2}{c|}{Absorption peaks} \\
				\hline
				Frequency & Reached value & Frequency & Reached value \\
				\hline
				303 Hz & 17.2dB & 315 Hz & 0.97\\
				\hline
				346 Hz & 16.9dB & 353 Hz & 0.95 \\
				\hline
				396 Hz & 16.2dB & 394 Hz & 0.96\\
				\hline
				1031 Hz & 13.3dB & 1030 Hz & 0.95\\
				\hline
				1151 Hz & 15.1dB & 1150 Hz & 0.96\\
				\hline
				1291 Hz & 14.4dB & 1295 Hz & 0.93\\								
				\hline
			\end{tabular}}}
		\end{table}

		\begin{figure}[!h]
			\centerline{\includegraphics[width=8.6cm,keepaspectratio]{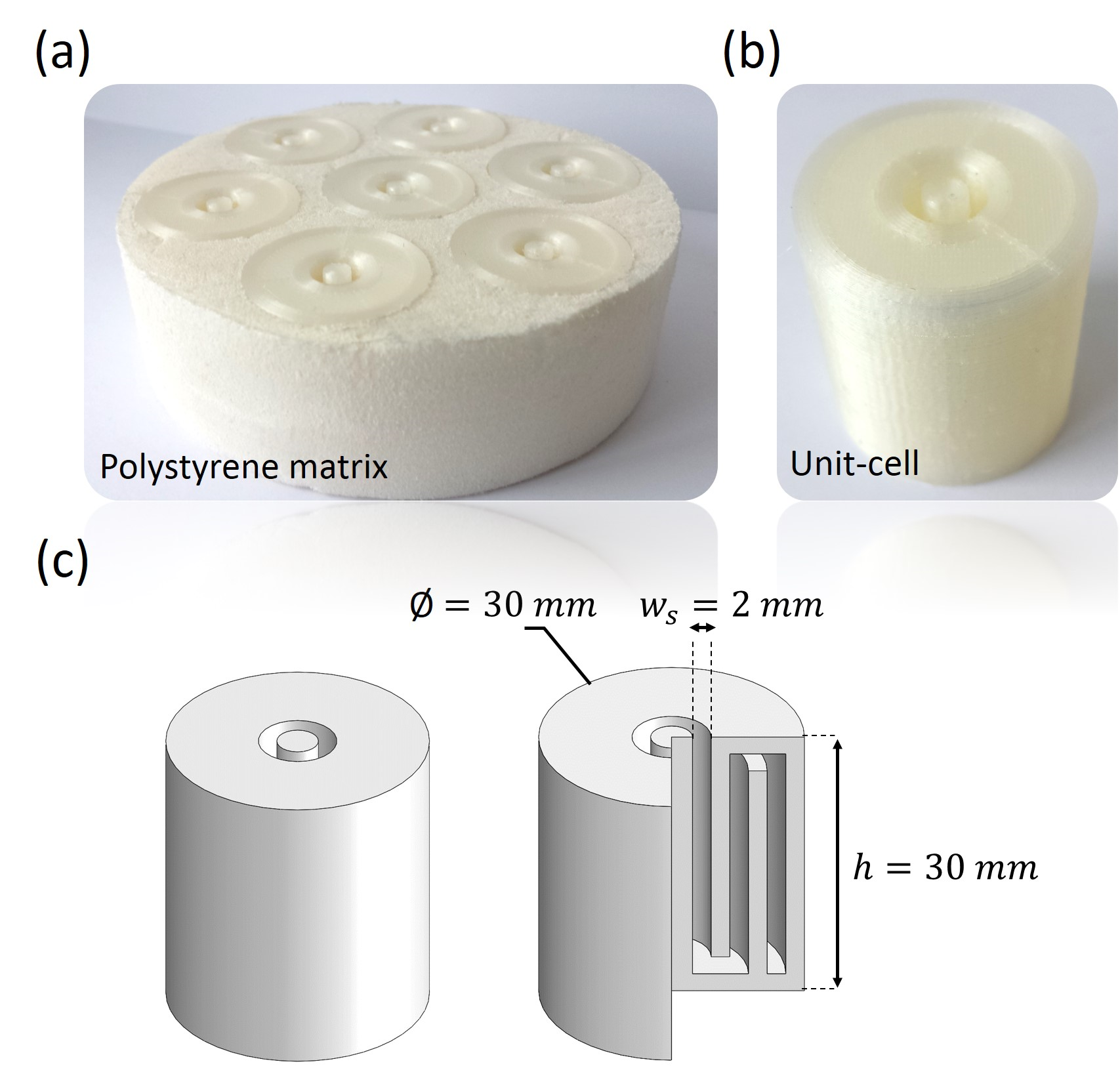}}
			\caption{Photograph of the sample constituted of a polystyrene matrix (a) in which are embedded several and identical metamaterial unit-cells (b), all fabricated by 3D printing. (c)Schematics of a unit-cell of acoustic metamaterial involving the space-coiled technique. The object is cut using two imaginary cutting planes. The unwanted portion is mentally discarded exposing the interior structuring. The diameter equals $30$ mm, the slit width have a value of $w_s=2$ mm, and the metamaterial height equals $h=30$ mm.}
			\label{fig:fig09}
		\end{figure}
		
\begin{figure}[!h]
			\centerline{\includegraphics[width=8.6cm,keepaspectratio]{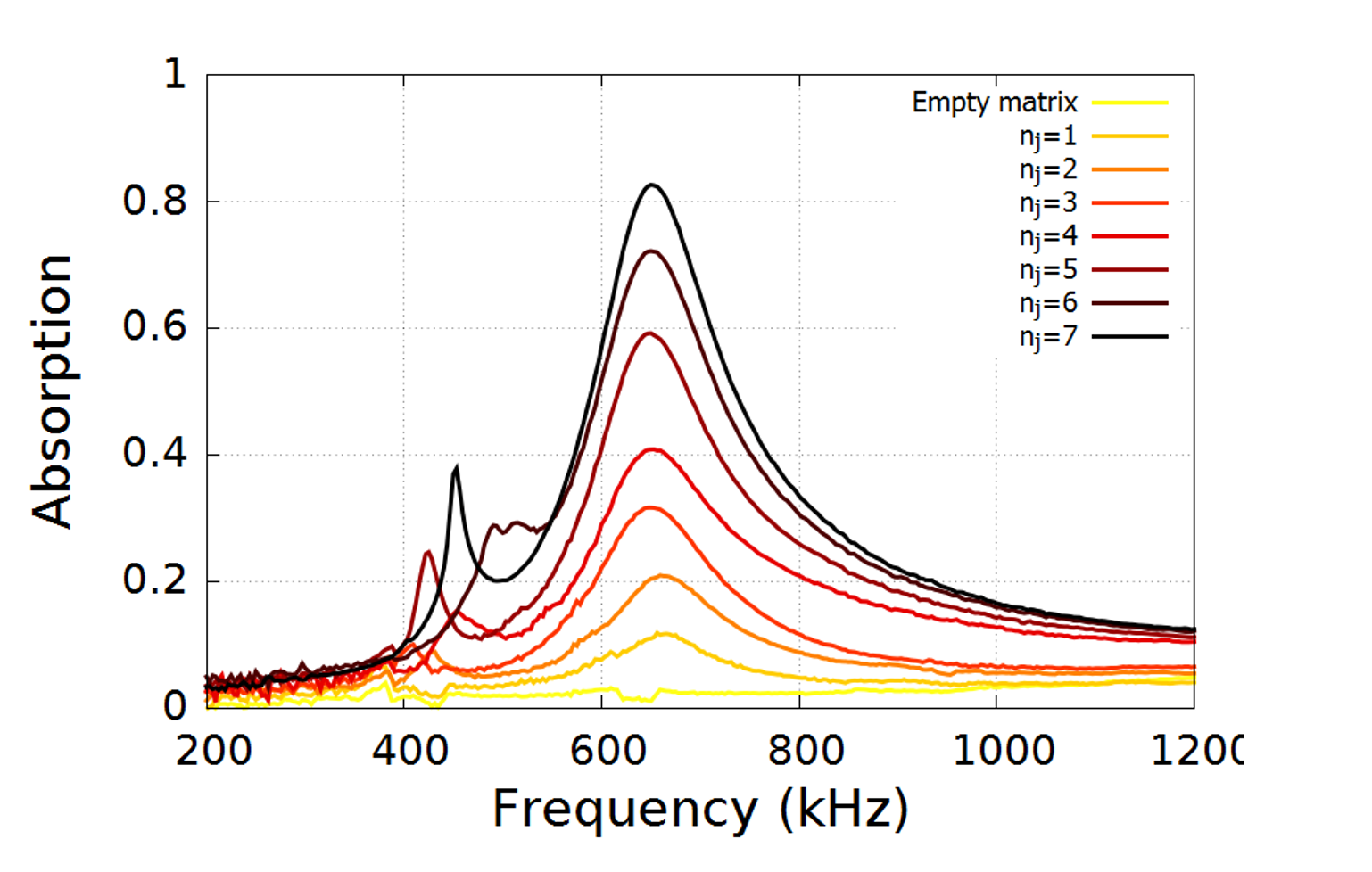}}
			\caption{Comparison of the absorption spectra related to the polystyrene matrix in which are embedded from $1$ unit-cell to $7$ unit-cells of acoustic metamatarial. As a reference, the absorption spectra corresponding to the polystyrene matrix is represented in yellow.}
			\label{fig:fig10}
		\end{figure}

		In what follows, we totally tailor the acoustic absorption using the filling ratio of the resonant cavities, which corresponds to the surface distribution of the resonators over the total section of the waveguide. To do this, we consider a cylindrical metamaterial unit-cell constituted of a space-coiled quarte-wave-like resonator having a length and a diameter of $30$ mm, but designed to have an effective length of around $130$ mm. The slit width is fixed to $w_s=2$ mm. We fabricate $7$ unit-cells as well as a reception matrix made of polystyrene, as can be seen in Figure~\ref{fig:fig09}a and b. We characterize various combinations to see how absorption evolves depending on the number of unit-cells used $n_j$, and represent it in Figure~\ref{fig:fig10}. For the configuration involving only one metamaterial unit-cell $n_j=1$, we clearly observe a peak of absorption with an amplitude of $0.1$. From there, it is interesting to see how the absorption increases, at the resonance frequency aimed, when adding from $1$ to $7$ unit-cells of acoustic metamatarial, until reaching $0.8$. As an element of comparison, we trace the absorption profile of the polystyrene matrix and see that taken alone, it does not generate any particular acoustic absorption of sound under $1$ kHz. Thus, given a fixed slit width, we can see that it is possible to modulate the level of sound absorption only by increasing the filling ratio of cross-sectional area of the resonator. We finally obtain an absorption peak reaching $0.83$, at the frequency of $650$ Hz, with a bandwidth of $30 \%$. This is achieved with a filling ratio of $3.64$ $\%$, and a metamaterial almost $\frac{\lambda}{h}=17$ smaller that the wavelength.

		The metamaterials presented in this paper thus provide an incredible amount of flexibility, both in terms of materials and performance. First of all, their subwavelength character is well suited to address miniaturization challenges in many application domains involving noise control. In addition, these metamaterials do not rely on the use of materials having intrinsic dissipative properties with regard to the sound. They are based on confining and raise the density of energy using resonances, which leads to enhance the visco-thermal effects of air, that would be neglected in free propagation. This gives the ability of functionalizing materials while taking advantage of their mechanical, optical properties, and more environmental friendly. Finally, the tailorability demonstrated in the present article is a powerful tool for applications for which the suppression of acoustic waves is not necessarily the challenge, but rather a finer control of acoustic properties in both frequency and amplitude. Added to this is the fact that our metamatarial allows the circulation of air flows and consequently thermal fluxes, societal impacts are certainly increased tenfold.

		\section{Conclusions} 
		
		In conclusion, we have experimentally demonstrated that controlling the confinement of the acoustic energy in a resonant cavity permits to enhance the dissipations due to visco-thermal effects, and therefore to tailor the level of sound absorption, even with low value of filling ratio. The experimental data are in good agreement with numerical results when visco-thermal effects are taken into account, which have been proved to be necessary when dealing with strong subwavelength structures made of channels and cavities. Moreover, we have presented an acoustic metamaterial constituted of multiple and neighboring resonant cavities and experimentally observed the generation of broadband absorption. Finally, we investigated the role of the filling ratio in tailoring such an absorption. To do this, we designed several and identical metamaterial unit-cells using space-coiled techniques, and embedded them in a polystyrene matrix. By doing so, we expect that such low density, low volume, flexibility and wide tailorability provided by these acoustic metamaterials make it suitable candidates for tremendous opportunities in soundproofing, but also to design more complex acoustic device for controlling wave propagation.
		
		\section*{Acknowledgements}
		
		The authors gratefully acknowledge Youssef Tejda for the fabrication of the device described in this work. The research leading to these results has received funding from the Region of Franche-Comte and financial support from the Labex ACTION program (Contact No. ANR-11-LABX-0001-01).

\bibliographystyle{apsrev4-1} 


%

\end{document}